\documentclass[a4paper]{report}
\usepackage[utf8]{inputenc}
\usepackage[T1]{fontenc}
\usepackage{RJournal}
\usepackage{amsmath,amssymb,array}
\usepackage{booktabs}

\usepackage{longtable}

\newlength{\cslhangindent}
\newlength{\csllabelwidth}
\newlength{\cslentryspacingunit} % times entry-spacing
  {}%
  {\par}
 {% don't indent paragraphs
  \setlength{\parindent}{0pt}
  \ifodd #1
  \let\oldpar\par
  \def\par{\hangindent=\cslhangindent\oldpar}
  \fi
  \setlength{\parskip}{#2\cslentryspacingunit}
 }%
 {}
\usepackage{calc}

\begin{document}

%% do not edit, for illustration only
\sectionhead{Contributed research article}
\volume{XX}
\volnumber{YY}
\year{20ZZ}
\month{AAAA}

\begin{article}
\renewcommand{\backref}[1]{}\pagestyle{plain}
% !TeX root = RJwrapper.tex
\title{A Unified Approach to Concurrent, Parallel Map-Reduce in R using Futures}

\author{by Henrik Bengtsson}

\maketitle

\abstract{%
The R ecosystem offers a rich variety of map-reduce application programming interfaces (APIs) for iterative computations, yet parallelizing code across these diverse frameworks requires learning multiple, often incompatible, parallel APIs. The \textbf{futurize} package addresses this challenge by providing a single function, futurize(), which transpiles sequential map-reduce expressions into their parallel equivalents in the future ecosystem, which performs all the heavy lifting. By leveraging R's native pipe operator, users can parallelize existing code with minimal refactoring -- often by simply appending `\textbar\textgreater{} futurize()' to an expression. The package supports classical map-reduce functions from base R, \textbf{purrr}, \textbf{crossmap}, \textbf{foreach}, \textbf{plyr}, \textbf{BiocParallel}, e.g., lapply(xs, fcn) \textbar\textgreater{} futurize() and map(xs, fcn) \textbar\textgreater{} futurize(), as well as a growing set of domain-specific packages, e.g., \textbf{boot}, \textbf{caret}, \textbf{glmnet}, \textbf{lme4}, \textbf{mgcv}, and \textbf{tm}. By abstracting away the underlying parallel machinery, and unifying handling of future options, the package enables developers to declare what to parallelize via futurize(), and end-users to choose how via plan(). This article describes the philosophy, design, and implementation of \textbf{futurize}, demonstrates its usage across various map-reduce paradigms, and discusses its role in simplifying parallel computing in R.
}

\hypertarget{sec-introduction}{%
\section{Introduction}\label{sec-introduction}}

A common data processing strategy in R \citep{RLanguage} is to use map-reduce calls, where a function is applied to each element of a collection. This map-reduce paradigm, also known as the split-apply-combine strategy \citep{Wickham_2011}, is implemented across several popular, yet different APIs. Base R provides \texttt{lapply()} and related functions, the \CRANpkg{purrr} package \citep{CRAN:purrr} offers functions like \texttt{map()}, and the \CRANpkg{foreach} package \citep{CRAN:foreach} introduces its own \texttt{foreach()\ \%do\%\ \{\ \}} construct. They all have their pros and cons, but ultimately they achieve the same end goal. When used correctly, they process each iteration concurrently\footnote{This article uses the term ``concurrent'' to describe the logical structure of computations and the term ``parallel'' to describe their physical execution on multiple resources.} and independently of the others and without side effects -- a property referred to as ``embarrassingly parallel'', which makes such computations well suited for concurrent, and parallel, execution.

Concurrent processing is essential for structuring computationally intensive tasks. As datasets grow larger and analyses become more complex, the ability to distribute work across multiple CPU cores or machines becomes increasingly valuable. The R ecosystem offers a rich variety of methods for concurrent, parallel processing, with many dating back more than a decade \citep{Schmidberger_etal_2009, Eddelbuettel_2021}. R gained built-in support for some via the \pkg{parallel} package in version 2.14.0 (2011), while additional, community-contributed parallel map-reduce APIs can be found on CRAN and Bioconductor. Some of the alternatives mimic the original API and behavior closely, whereas others only loosely so. For example, when using \texttt{parallel::mclapply()}, in contrast to \texttt{parallel::lapply()}, errors are internally caught via \texttt{try()} and, instead of being signaled, are returned as \texttt{try-error} conditions, with no reference to the original error object. Likewise, the \texttt{parLapply()} function catches errors and does not give access to the original error object. On the other hand, in contrast to \texttt{mclapply()}, it does signal an error, with a message string containing the original error message. Neither of them propagates messages or warnings, which are completely lost, as is standard output (``stdout''). From looking not only at the source code of CRAN and Bioconductor packages using these functions, but also at online support forums, it is clear that many developers and end-users assume these functions behave like \texttt{lapply()} and they are not aware of the above differences. To the author's knowledge, the future ecosystem \citep{Bengtsson_2021} is the only parallelization framework in R that preserves errors and relays output and conditions out of the box.

Continuing, many parallel map-reduce functions have their own sub-APIs for fine-tuning how parallelization is executed, e.g., the number of parallel processes and of what type. This fragmentation not only adds unnecessary mental overhead for the developer; it also presents a lock-in challenge. The developer often has to spend a significant amount of time assessing alternatives before choosing a specific parallel framework to support, which increases the risk of lock-in afterward due to a reluctance to evaluate another solution. The lock-in impacts the developer's coding style options going forward, but it also risks limiting the end-users' options on how and where to execute the code. For example, when using \texttt{mclapply()}, parallelization is limited to Unix and macOS, leaving out users on Windows. Similarly, using \texttt{parLapply()} with \texttt{makeCluster()}, or \texttt{foreach()\ \%dopar\%\ \{\ \}} with common \CRANpkg{doParallel} \citep{CRAN:doParallel} settings, limits parallelization to the local computer. The central design philosophy of the future ecosystem is to minimize such developer and end-user lock-in and instead allow implementations to automatically remain agile to future technology improvements. The \CRANpkg{futurize} package \citep{CRAN:futurize}, presented here, is part of this core design strategy.

The concept of futures for parallel evaluation was introduced in the late 1970s \citep{Hibbard_1976, HewittBaker_1977, FriedmanWise_1978}. In R, the \CRANpkg{future} package \citep{CRAN:future} established a unifying framework for parallel and distributed processing using this concept.
Building on this foundation, several packages emerged to provide parallel versions of common map-reduce APIs, e.g., \CRANpkg{future.apply} \citep{CRAN:future.apply} for base R's apply functions, \CRANpkg{furrr} \citep{CRAN:furrr} for \CRANpkg{purrr}'s alternatives, and \CRANpkg{doFuture} \citep{CRAN:doFuture} for \CRANpkg{foreach}'s iteration constructs. While these future-based extensions successfully bring parallelization to their respective APIs, they require developers to learn and use slightly different function names and parallel-option conventions, e.g., parallelizing \texttt{lapply()} requires switching to \texttt{future.apply::future\_lapply()}, parallelizing \texttt{purrr::map()} requires switching to \texttt{furrr::future\_map()}, and parallelizing \texttt{foreach::foreach()\ \%do\%\ \{\ \}} requires switching to \texttt{foreach::foreach()\ \%dofuture\%\ \{\ \}} of \CRANpkg{doFuture}. Beyond the different function names, each package has its own conventions for specifying options on how to fine-tune parallel execution. Although small, these differences are enough to require different documentation across packages, which in turn fragments the developer and user bases that miss out on cross-pollination that is otherwise available from working with a more unified API.

The \CRANpkg{futurize} package addresses these challenges by providing a single function, \texttt{futurize()}, that can parallelize any supported sequential map-reduce expression. Rather than requiring developers to replace function calls with custom parallel counterparts, \CRANpkg{futurize} allows them to keep their familiar sequential code and simply pipe it to \texttt{futurize()};

\begin{verbatim}
library(futurize)
plan(multisession)

xs <- 1:100
ys <- lapply(xs, fcn) |> futurize()
\end{verbatim}

This approach offers several advantages:

\begin{enumerate}
\def\labelenumi{\arabic{enumi}.}
\item
  \textbf{Minimal code changes}: Parallelization requires adding only \texttt{\textbar{}\textgreater{}\ futurize()} to existing expressions.
\item
  \textbf{Familiar code}: The original sequential logic remains visible and readable, keeping the focus on the map-reduce logic away from parallelization details.
\item
  \textbf{Familiar behavior}: Standard output and conditions -- including messages, warnings, and errors -- behave the same as with the original sequential logic.
\item
  \textbf{Unified interface}: The same \texttt{futurize()} function works across all supported map-reduce APIs.
\item
  \textbf{Unified options}: The \texttt{futurize()} function provides a single, consistent way to specify parallel options regardless of the underlying API.
\item
  \textbf{Backend independence}: As with all future ecosystem tools, the choice of parallel backend is left to the end-user, e.g., \texttt{plan(multisession)}, \texttt{plan(future.mirai::mirai\_multisession)}, and \texttt{plan(future.batchtools::batchtools\_slurm)}.
\end{enumerate}

This article describes the philosophy, design, and implementation of \CRANpkg{futurize}. Section 2 discusses the philosophy and design principles underlying the package. Section 3 details the implementation, including the transpilation mechanism and supported APIs. Section 4 presents results demonstrating the package's usage across various map-reduce paradigms. Section 5 discusses the package's role in the future ecosystem and future development directions.

\hypertarget{sec-philosophy}{%
\section{Philosophy and design of futurize}\label{sec-philosophy}}

The design of \CRANpkg{futurize} is a direct extension of the core philosophy of the \CRANpkg{future} framework: to cleanly separate the developer's concern (\emph{what} to parallelize) from the end-user's concern (\emph{how} and \emph{where} to run it). While the \CRANpkg{future} package provides the low-level API (\texttt{future()}, \texttt{resolved()}, and \texttt{value()}) to achieve this, its direct application can be verbose for common map-reduce tasks. Well-established, higher-level APIs like \CRANpkg{future.apply} and \CRANpkg{furrr} provide more convenient functions, which hide away parts of the parallelization orchestration, but still require developers to learn and use new function names compared to their original, sequential counterparts.

The \CRANpkg{futurize} package bridges this gap by acting as a high-level ``transpiler.'' The central idea is that developers should be able to write their code using the familiar, sequential map-reduce functions they already know, and then ``opt-in'' to parallelization with a single, simple function.
The goal is to minimize the mental load and the amount of mental context switching required when turning sequential code into parallel code.

\hypertarget{design-principles}{%
\subsection{Design principles}\label{design-principles}}

The design of \CRANpkg{futurize} is guided by several core principles inherited from the broader future ecosystem philosophy, along with additional principles specific to its role as a transpilation layer.

\textbf{Separation of concerns.} Following the future ecosystem philosophy, \CRANpkg{futurize} maintains a strict separation between what to parallelize (concurrently) and how to parallelize. Developers and users specify what to parallelize by piping expressions to \texttt{futurize()}. The choice of parallel backend (e.g., local multi-CPU-core, cluster, cloud, or high-performance compute (HPC) scheduler) is entirely controlled by the end-user through \texttt{plan()}. This separation ensures that code written with \CRANpkg{futurize} automatically works on any current or future parallel backend \footnote{In order to guarantee that code using futures works with any future backend, future backends must be compliant with the Future API. Compliance is validated using the \CRANpkg{future.tests} package \citep{CRAN:future.tests}.}.

\textbf{Minimal API surface.} The package exposes primarily a single function, \texttt{futurize()}, which handles all transpilation. This minimalist design reduces the learning curve and makes the package easy to adopt. Options for controlling future resolution are passed as arguments to \texttt{futurize()} itself.

\textbf{Preserve existing code.} Unlike approaches that require rewriting sequential code to use parallel functions, \CRANpkg{futurize} allows the original sequential logic to remain intact. The expression \texttt{lapply(xs,\ fcn)\ \textbar{}\textgreater{}\ futurize()} clearly shows what operation is being performed (\texttt{lapply()}) and that it can be parallelized (\texttt{futurize()}). This transparency aids code review and maintenance.

\textbf{Global disable/enable.} For debugging and profiling purposes, \CRANpkg{futurize} can be globally disabled by calling \texttt{futurize(FALSE)}. All \texttt{futurize()} calls then pass through as if \texttt{\textbar{}\textgreater{}\ futurize()} does not exist. This is reenabled with \texttt{futurize(TRUE)}. Importantly, only end-users may toggle this setting; packages must never globally disable or enable futurization.

\hypertarget{the-transpilation-approach}{%
\subsection{The transpilation approach}\label{the-transpilation-approach}}

The term ``transpilation'' describes the process of transforming source code from one form into another, a.k.a. source-to-source translation \citep{Loveman_1977, Aho_etal_2006}. In \CRANpkg{futurize}, transpilation transforms sequential map-reduce expressions into logically equivalent expressions that leverage the \CRANpkg{future} ecosystem for concurrent execution. When \texttt{futurize()} receives an expression such as \texttt{lapply(xs,\ fcn)}, it captures the unevaluated call and identifies both the map-reduce call and its main function. Based on this information, \CRANpkg{futurize} rewrites the expression into its future-based counterpart. For example,

\begin{itemize}
\item
  \texttt{base::lapply()\ \textbar{}\textgreater{}\ futurize()} transpiles to \texttt{future.apply::future\_lapply()}
\item
  \texttt{purrr::map()\ \textbar{}\textgreater{}\ futurize()} transpiles to \texttt{furrr::future\_map()}, and
\item
  \texttt{foreach::foreach()\ \%do\%\ \{\ \}\ \textbar{}\textgreater{}\ futurize()} transpiles to \texttt{foreach::foreach()\ \%dofuture\%\ \{\ \}}, where \texttt{\%dofuture\%} is part of \CRANpkg{doFuture}.
\end{itemize}

This approach is more flexible than simple function replacement because the transpilation can handle the nuances of different APIs, including differences in argument naming and default values. Parallelization options (a.k.a. future options) specified via \texttt{futurize()} are mapped appropriately to the underlying implementation, allowing developers to work with a single, unified interface regardless of which map-reduce API is being used.

Importantly, the transpilation approach decouples user-facing semantics from implementation details. In addition, this separation enables transparent evolution of the underlying implementation. For example, although futurization of \texttt{lapply()} currently relies on \CRANpkg{future.apply}, future versions of \CRANpkg{futurize} may transpile directly into calls using only the core \CRANpkg{future} API, without affecting existing user code or changing how \CRANpkg{futurize} is used.

By combining source-to-source translation with domain-specific language (DSL) principles \citep{Fowler_2010}, \CRANpkg{futurize} provides a minimalistic, expressive syntax for introducing concurrency into existing R code while preserving readability, composability, and long-term stability.

\hypertarget{integration-with-rs-pipe-operator}{%
\subsection{Integration with R's pipe operator}\label{integration-with-rs-pipe-operator}}

A key design decision in \CRANpkg{futurize} is its integration with R's native pipe operator, \texttt{\textbar{}\textgreater{}}, introduced in R 4.1.0. The pipe enables natural expression of the ``evaluate this, but concurrently'' intent:

\begin{verbatim}
ys <- lapply(xs, fcn) |> futurize()
\end{verbatim}

This syntax places the parallelization modifier at the end of the expression, keeping the core logic prominent. By definition of the native pipe operator, \texttt{expr\ \textbar{}\textgreater{}\ futurize()} is identical to \texttt{futurize(expr)}, meaning the above example could also be written as \texttt{ys\ \textless{}-\ futurize(lapply(xs,\ fcn))}, but the pipe syntax is preferred for its readability.

\hypertarget{unified-future-options}{%
\subsection{Unified future options}\label{unified-future-options}}

One of the main objectives of \CRANpkg{futurize} is to provide a unified way to specify parallel options through the \texttt{futurize()} function. The underlying parallel packages, \CRANpkg{future.apply}, \CRANpkg{furrr}, and \CRANpkg{doFuture}, each have their own, slightly different, conventions for specifying future options such as random-number generation and global variable handling, but also for controlling load balancing. For example, \CRANpkg{future.apply} uses arguments like \texttt{future.seed}, \texttt{future.scheduling}, while \CRANpkg{furrr} uses \texttt{.options} with \texttt{furrr\_options()}, and \CRANpkg{doFuture}'s \texttt{\%dofuture\%} uses \texttt{foreach()} argument \texttt{.options.future}. These differences create friction when switching between APIs, or unnecessary complexity in documenting them and for users trying to understand their differences.
The \texttt{futurize()} function abstracts away these differences, providing a single, consistent interface:

\begin{verbatim}
ys <- lapply(xs, fcn) |> futurize(seed = TRUE, chunk_size = 2)
\end{verbatim}

Key options include:

\begin{itemize}
\item
  \texttt{seed}: Controls parallel random-number generation. When \texttt{TRUE}, L'Ecuyer-CMRG streams \citep{LEcuyer_1999, LEcuyer_etal_2017} are used to ensure reproducible, statistically sound random numbers across parallel workers. For some functions, such as \texttt{replicate()} and \texttt{foreach::times()}, \texttt{futurize()} defaults to \texttt{seed\ =\ TRUE}.
\item
  \texttt{globals}: Controls identification and export of global variables. By default, globals are automatically identified through static-code analysis \citep{CRAN:globals, CRAN:codetools}.
\item
  \texttt{packages}: Specifies packages to attach on workers. By default, packages are automatically inferred from the globals identified.
\item
  \texttt{stdout} and \texttt{conditions}: Control capture and relay of standard output, messages, and warnings from workers.
\item
  \texttt{scheduling} and \texttt{chunk\_size}: Control load balancing by determining how many elements are processed per future.
\end{itemize}

Regardless of whether the underlying expression uses \texttt{lapply()}, \texttt{purrr::map()}, etc., the same future options work consistently. This unification significantly simplifies the mental model for users, particularly for those working with multiple map-reduce APIs. The unification also makes it easy to add further options at a later point in time.

\hypertarget{sec-implementation}{%
\section{Implementation}\label{sec-implementation}}

\hypertarget{package-dependencies-and-requirements}{%
\subsection{Package dependencies and requirements}\label{package-dependencies-and-requirements}}

The \CRANpkg{futurize} package requires R version 4.1.0 (May 2021) or later to support the native pipe operator\footnote{Theoretically, \CRANpkg{futurize} works with older versions of R than 4.1.0, by using \texttt{futurize(expr)}. However, in order to release \CRANpkg{futurize} on CRAN with \texttt{\textbar{}\textgreater{}} in the documentation and its examples, it needs to depend on R 4.1.0.}. Although strictly not necessary, it depends on \CRANpkg{future} for practical purposes, such as making \texttt{plan()} available after \texttt{library(futurize)}. All other package dependencies are optional (``suggested'') in the sense that they are only required if needed to futurize the map-reduce call of interest. For example, futurization of base-R map-reduce calls requires \CRANpkg{future.apply}, \CRANpkg{purrr} calls require \CRANpkg{furrr}, and \CRANpkg{foreach} calls require \CRANpkg{doFuture}. See Table 1 for currently supported functions and their package dependencies. These ``buy-in'' package dependencies are loaded lazily; developers and users only need to install the packages relevant to their use case. Package developers need to import these packages if used - importing them asserts that they are installed as part of the package using them via \texttt{futurize()}.

\hypertarget{implementation-of-the-transpilation-process}{%
\subsection{Implementation of the transpilation process}\label{implementation-of-the-transpilation-process}}

This section describes how \CRANpkg{futurize} implements the transpilation approach outlined in Section 2.2. It performs this in a series of steps:

\begin{enumerate}
\def\labelenumi{\arabic{enumi}.}
\item
  \textbf{Expression capture}: When an expression is piped into \texttt{futurize()}, the function uses \texttt{substitute()} to capture the unevaluated call. This is a well-established non-standard evaluation (NSE) \citep{Chambers_2008, Wickham_2019} technique in R that allows functions to operate on code as data.
\item
  \textbf{Function identification}: It then inspects the call to identify the function being invoked. It determines both the function name and its originating namespace, which is critical for distinguishing functions with the same name in different packages.
\item
  \textbf{Transpiler lookup}: The package maintains an internal registry of transpilers for known functions. It queries this registry for a transpiler that matches the identified function.
\item
  \textbf{Expression rewriting}: Once the correct transpiler is found, it rewrites the original expression. The transpiler maps arguments from the original call to the corresponding arguments in the target function and passes along parallelization options specified via \texttt{futurize()}.
\item
  \textbf{Evaluation}: Finally, the newly formed expression is evaluated in the original environment (the parent frame).
\end{enumerate}

To see how different calls are transpiled, use \texttt{futurize(eval\ =\ FALSE)}. This returns the transpiled call without evaluating it.

\hypertarget{sec-unwrap}{%
\subsection{Expression unwrapping}\label{sec-unwrap}}

For practical purposes, the transpilation mechanism includes logic to ``unwrap'' expressions enclosed in constructs such as \texttt{\{\ \}}, \texttt{(\ )}, \texttt{local()}, \texttt{I()}, \texttt{identity()}, \texttt{suppressMessages()}, and \texttt{suppressWarnings()}. The transpiler descends through wrapping constructs until it finds a transpilable expression, avoiding the need to place \texttt{futurize()} inside such constructs. This allows for patterns like:

\begin{verbatim}
ys <- {
  lapply(xs, fcn)
} |> suppressMessages() |> futurize()
\end{verbatim}

avoiding having to write:

\begin{verbatim}
ys <- {
  lapply(xs, fcn) |> futurize()
} |> suppressMessages()
\end{verbatim}

A more complex example on unwrapping is given in Section 4.10 when discussing progress reporting.

\hypertarget{supported-apis}{%
\subsection{Supported APIs}\label{supported-apis}}

The \CRANpkg{futurize} package supports transpilation of functions from multiple packages. Tables 1 and 2 summarize the supported map-reduce and domain-specific functions, respectively.

\begin{longtable}[]{@{}
  >{\raggedright\arraybackslash}p{(\columnwidth - 4\tabcolsep) * \real{0.1288}}
  >{\raggedright\arraybackslash}p{(\columnwidth - 4\tabcolsep) * \real{0.7117}}
  >{\raggedright\arraybackslash}p{(\columnwidth - 4\tabcolsep) * \real{0.1595}}@{}}
\caption{Map-reduce functions currently supported by \texttt{futurize()} for parallel transpilation.}\tabularnewline
\toprule\noalign{}
\begin{minipage}[b]{\linewidth}\raggedright
Package
\end{minipage} & \begin{minipage}[b]{\linewidth}\raggedright
Functions
\end{minipage} & \begin{minipage}[b]{\linewidth}\raggedright
Requires
\end{minipage} \\
\midrule\noalign{}
\endfirsthead
\toprule\noalign{}
\begin{minipage}[b]{\linewidth}\raggedright
Package
\end{minipage} & \begin{minipage}[b]{\linewidth}\raggedright
Functions
\end{minipage} & \begin{minipage}[b]{\linewidth}\raggedright
Requires
\end{minipage} \\
\midrule\noalign{}
\endhead
\bottomrule\noalign{}
\endlastfoot
\pkg{base} & \texttt{lapply()}, \texttt{sapply()}, \texttt{tapply()}, \texttt{vapply()}, \texttt{mapply()}, \texttt{.mapply()}, \texttt{Map()}, \texttt{eapply()}, \texttt{apply()}, \texttt{by()}, \texttt{replicate()}, \texttt{Filter()} & \CRANpkg{future.apply} \\
\pkg{stats} & \texttt{kernapply()} & \CRANpkg{future.apply} \\
\CRANpkg{purrr} & \texttt{map()} and variants, \texttt{map2()} and variants, \texttt{pmap()} and variants, \texttt{imap()} and variants, \texttt{modify()}, \texttt{modify\_if()}, \texttt{modify\_at()}, \texttt{map\_if()}, \texttt{map\_at()}, \texttt{invoke\_map()} & \CRANpkg{furrr} \\
\CRANpkg{crossmap} & \texttt{xmap()} and variants, \texttt{xwalk()}, \texttt{map\_vec()}, \texttt{map2\_vec()}, \texttt{pmap\_vec()}, \texttt{imap\_vec()} & (itself) \\
\CRANpkg{foreach} & \texttt{\%do\%}, e.g., \texttt{foreach()\ \%do\%\ \{\ \}}, \texttt{times()\ \%do\%\ \{\ \}} & \CRANpkg{doFuture} \\
\CRANpkg{plyr} & \texttt{aaply()} and variants, \texttt{ddply()} and variants, \texttt{llply()} and variants, \texttt{mlply()} and variants & \CRANpkg{doFuture} \\
\BIOpkg{BiocParallel} & \texttt{bplapply()}, \texttt{bpmapply()}, \texttt{bpvec()}, \texttt{bpiterate()}, \texttt{bpaggregate()} & \CRANpkg{doFuture} \\
\end{longtable}

\begin{longtable}[]{@{}
  >{\raggedright\arraybackslash}p{(\columnwidth - 4\tabcolsep) * \real{0.1593}}
  >{\raggedright\arraybackslash}p{(\columnwidth - 4\tabcolsep) * \real{0.6637}}
  >{\raggedright\arraybackslash}p{(\columnwidth - 4\tabcolsep) * \real{0.1770}}@{}}
\caption{Domain-specific functions currently supported by \texttt{futurize()} for parallel transpilation.}\tabularnewline
\toprule\noalign{}
\begin{minipage}[b]{\linewidth}\raggedright
Package
\end{minipage} & \begin{minipage}[b]{\linewidth}\raggedright
Functions
\end{minipage} & \begin{minipage}[b]{\linewidth}\raggedright
Requires
\end{minipage} \\
\midrule\noalign{}
\endfirsthead
\toprule\noalign{}
\begin{minipage}[b]{\linewidth}\raggedright
Package
\end{minipage} & \begin{minipage}[b]{\linewidth}\raggedright
Functions
\end{minipage} & \begin{minipage}[b]{\linewidth}\raggedright
Requires
\end{minipage} \\
\midrule\noalign{}
\endhead
\bottomrule\noalign{}
\endlastfoot
\CRANpkg{boot} & \texttt{boot()}, \texttt{censboot()}, \texttt{tsboot()} & \CRANpkg{future} \\
\CRANpkg{caret} & \texttt{bag()}, \texttt{gafs()}, \texttt{nearZeroVar()}, \texttt{rfe()}, \texttt{safs()}, \texttt{sbf()}, \texttt{train()} & \CRANpkg{doFuture} \\
\CRANpkg{glmnet} & \texttt{cv.glmnet()} & \CRANpkg{doFuture} \\
\CRANpkg{lme4} & \texttt{allFit()}, \texttt{bootMer()} & \CRANpkg{future} \\
\CRANpkg{mgcv} & \texttt{bam()}, \texttt{predict.bam()} & \CRANpkg{future} \\
\CRANpkg{tm} & \texttt{TermDocumentMatrix()}, \texttt{tm\_index()}, \texttt{tm\_map()} & \CRANpkg{future} \\
\end{longtable}

To see which packages are currently supported, use:

\begin{verbatim}
futurize_supported_packages()
#>  [1] "base"         "BiocParallel" "boot"         "caret"        "crossmap"    
#>  [6] "foreach"      "glmnet"       "lme4"         "mgcv"         "plyr"        
#> [11] "purrr"        "stats"        "tm"
\end{verbatim}

To see which functions are supported for a specific package, use:

\begin{verbatim}
futurize_supported_functions("caret")
#> [1] "bag"         "gafs"        "nearZeroVar" "rfe"         "safs"       
#> [6] "sbf"         "train"
\end{verbatim}

\hypertarget{sec-results}{%
\section{Results}\label{sec-results}}

This section demonstrates \CRANpkg{futurize} usage across various map-reduce and domain-specific paradigms, highlighting its unified approach to parallelization.
The map-reduce examples cover base R, \CRANpkg{purrr}, \CRANpkg{foreach}, and \CRANpkg{plyr} \citep{CRAN:plyr}.
The domain-specific examples cover packages \CRANpkg{boot} \citep{CRAN:boot}, \CRANpkg{lme4} \citep{CRAN:lme4}, \CRANpkg{glmnet} \citep{CRAN:glmnet}, and \CRANpkg{caret} \citep{CRAN:caret}.
Most examples are overly simplified for the purpose of readability while still illustrating how \texttt{futurize()} applies to different types of map-reduce calls. They are not meant to illustrate real-world use cases.
For all examples, we assume a parallel backend has been registered by the user:

\begin{verbatim}
library(future)
plan(multisession)
\end{verbatim}

This particular setup uses parallel workers on the local machine leveraging PSOCK parallelization of the \pkg{parallel} package. As a matter of fact, any compliant backend\footnote{See \url{https://www.futureverse.org/backends.html} for compliant future backends.} supported by the \CRANpkg{future} framework can be used in these examples without changing their code, including \CRANpkg{future.callr} \citep{CRAN:future.callr, CRAN:callr}, \CRANpkg{future.mirai} \citep{CRAN:future.mirai, CRAN:mirai}, and HPC schedulers via \CRANpkg{future.batchtools} \citep{CRAN:future.batchtools, Lang_etal_2017}.

\hypertarget{basic-usage-with-base-r}{%
\subsection{\texorpdfstring{Basic usage with base R }{Basic usage with base R }}\label{basic-usage-with-base-r}}

A common use case is to parallelize \texttt{lapply()} calls. For example, consider a function that performs some slow computation:

\begin{verbatim}
slow_fcn <- function(x) {
  Sys.sleep(1.0)  # Simulate work
  x^2
}
\end{verbatim}

When we apply this function to 100 elements using:

\begin{verbatim}
xs <- 1:100
ys <- lapply(xs, slow_fcn)
\end{verbatim}

it takes \textasciitilde100 seconds to complete. After having identified this bottleneck in the pipeline, and exhausted options for increasing the performance of \texttt{slow\_fcn()} itself, it is natural to turn to solutions that evaluate \texttt{slow\_fcn(xs{[}1{]})}, \ldots, \texttt{slow\_fcn(xs{[}100{]})} concurrently, in order to decrease the overall turnaround time (``walltime''). With the above local `multisession' workers, we can achieve this with \CRANpkg{futurize} using:

\begin{verbatim}
xs <- 1:100
ys <- lapply(xs, slow_fcn) |> futurize()
\end{verbatim}

The only change to the code is the appending of \texttt{\textbar{}\textgreater{}\ futurize()}. The logic of the computation remains unchanged. Figure 1 provides an overview how map-reduce is evaluated without and with \texttt{futurize()} processing.

\begin{figure}
\includegraphics[width=1\linewidth,alt={Diagram comparing sequential and parallel execution in R. On the left, a “MAIN R SESSION” runs xs <- 1:8 and ys <- lapply(xs, fcn), calling fcn(xs[1]) through fcn(xs[8]) one after another in the same session, then collecting results into ys <- list(fcn[1], ..., fcn[8]). On the right, the same code with |> futureize() distributes the calls across three workers: Worker #1 handles xs[1–3], Worker #2 handles xs[4–6], Worker #3 handles xs[7–8] (then idle). Results flow back to the main session, which assembles ys from all function outputs.}]{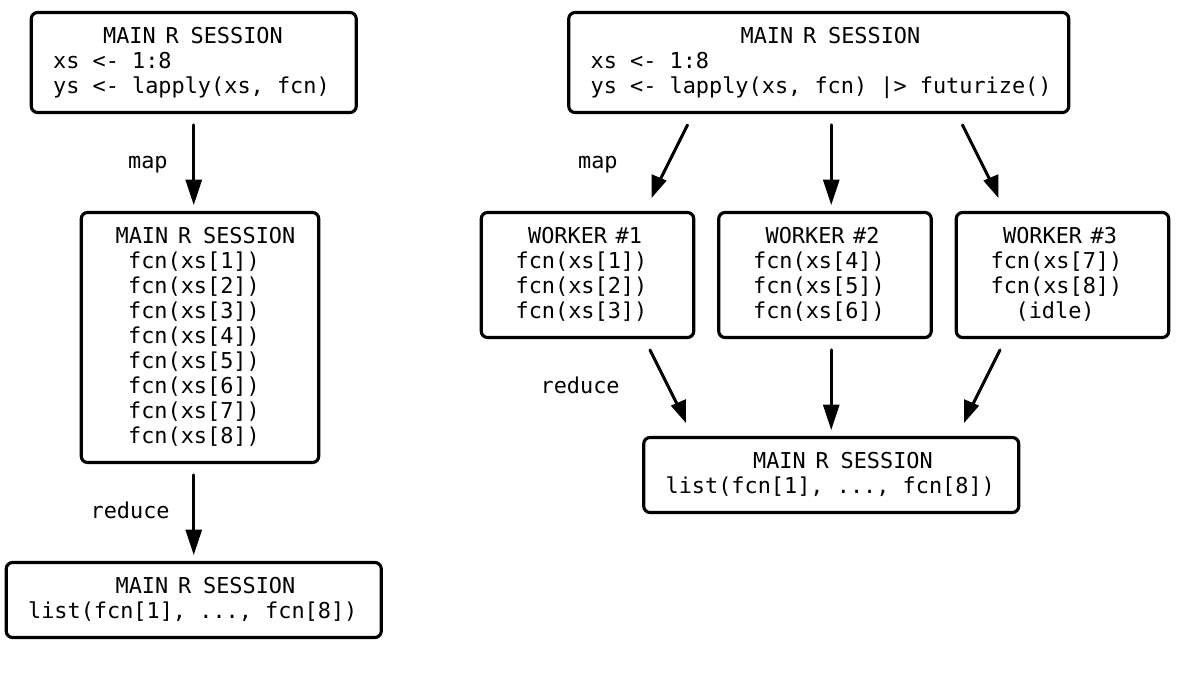} \caption{An illustration of how a map-reduce \texttt{lapply()} call is evaluated without (left) and with \texttt{futurize()} (right). There are in total eight \texttt{fcn()} calls - one per element in input vector \texttt{xs}. By default, these function calls, or ``tasks'', are evaluated sequentially one after the other. With \texttt{futurize()}, the evaluation is handled by the future ecosystem. For example, with three parallel workers, the tasks are distributed to workers and processed concurrently, after which their results are collected and reduced to its final form. Depending on future \texttt{plan()} set, parallel workers may run locally on the current machine or remotely on external machines. When futurizing \texttt{purrr::map()} and \texttt{foreach::foreach()\ \%do\%\ \{\ \}}, the parallel orchestration works the same way.}\label{fig:unnamed-chunk-1}
\end{figure}

In addition to other traditional map-reduce functions such as \texttt{vapply()} and \texttt{tapply()}, there are also functions like \texttt{replicate()}, which can be used to evaluate the same R expression multiple times. It can be parallelized as:

\begin{verbatim}
samples <- replicate(100, rnorm(10)) |> futurize()
\end{verbatim}

Because \texttt{replicate()} is predominantly used for resampling purposes (e.g., permutation tests), which requires RNG, \texttt{futurize()} defaults to \texttt{futurize(seed\ =\ TRUE)} when it detects that \texttt{replicate()} is used. This ensures that statistically sound, reproducible, and independent random numbers are produced regardless of which future backend is used \citep{Bengtsson_2021}.

\hypertarget{usage-with-purrr}{%
\subsection{\texorpdfstring{Usage with `purrr' }{Usage with `purrr' }}\label{usage-with-purrr}}

The \CRANpkg{purrr} package provides a popular functional programming toolkit. Its \texttt{map()} family of functions, which is closely related to the base R counterpart, parallelizes seamlessly with \CRANpkg{futurize}, e.g.

\begin{verbatim}
library(purrr)

xs <- 1:100

# Sequential
ys <- map(xs, slow_fcn)

# Parallel
ys <- map(xs, slow_fcn) |> futurize()
\end{verbatim}

More complex pipelines work naturally. For example, in:

\begin{verbatim}
ys <- 1:100 |>
  map(rnorm, n = 10) |> futurize(seed = TRUE) |>
  map_dbl(mean) |> futurize()
\end{verbatim}

both ``map'' calls are parallelized, with proper random-number handling for the first one.

\hypertarget{usage-with-foreach}{%
\subsection{\texorpdfstring{Usage with `foreach' }{Usage with `foreach' }}\label{usage-with-foreach}}

The \CRANpkg{foreach} package provides an alternative iteration paradigm, which was popularized thanks to its optional support for processing iterations in parallel via different ``foreach adapters''. With \CRANpkg{futurize}, the \texttt{\%do\%} operator can be parallelized directly:

\begin{verbatim}
library(foreach)

xs <- 1:100

# Sequential
ys <- foreach(x = xs) %do% { slow_fcn(x) }

# Parallel
ys <- foreach(x = xs) %do% { slow_fcn(x) } |> futurize()
\end{verbatim}

Despite using \CRANpkg{foreach}, there is no need for changing \texttt{\%do\%} to \texttt{\%dopar\%} or \texttt{\%dofuture\%}, or registering a foreach adapter -- that is all abstracted away by the \texttt{futurize()} transpiler.

Continuing, the \texttt{times()} function, which resembles \texttt{replicate()} in base R, works as well, e.g.

\begin{verbatim}
samples <- times(100) %do% rnorm(10) |> futurize()
\end{verbatim}

For the same reasons as when \texttt{replicate()} is used, \texttt{futurize()} defaults to \texttt{futurize(seed\ =\ TRUE)} when \texttt{times()} is used.

In addition, iterators of the \CRANpkg{iterators} package \citep{CRAN:iterators}, tightly coupled with the \CRANpkg{foreach} package, also work as expected, e.g.

\begin{verbatim}
library(iterators)

df <- data.frame(a = 1:4, b = letters[1:4])

ys <- foreach(d = df, i = icount()) %do% {
  list(value = d, index = i)
} |> futurize()
\end{verbatim}

\hypertarget{usage-with-plyr}{%
\subsection{\texorpdfstring{Usage with `plyr' }{Usage with `plyr' }}\label{usage-with-plyr}}

The \CRANpkg{plyr} package provides a large set of map-reduce functions. Despite being officially retired, the package is still maintained and widely used. Its \texttt{llply()} function, which closely resembles \texttt{lapply()} and \texttt{purrr:map()}, can be parallelized as:

\begin{verbatim}
library(plyr)

xs <- 1:100

# Sequential
ys <- llply(xs, slow_fcn)

# Parallel
ys <- llply(xs, slow_fcn) |> futurize()
\end{verbatim}

\hypertarget{other-map-reduce-packages}{%
\subsection{Other map-reduce packages}\label{other-map-reduce-packages}}

The \CRANpkg{crossmap} package \citep{CRAN:crossmap} adds to the \CRANpkg{purrr}-set of functions. For example, \texttt{xmap()} can apply a function to every combination of elements in a list. Using \CRANpkg{futurize}, its functions can be executed in parallel.
For users of Bioconductor, \CRANpkg{futurize} provides seamless integration with \BIOpkg{BiocParallel} \citep{BIOC:BiocParallel} -- one of the core Bioconductor packages. This allows Bioconductor workflows to leverage all parallel backends available in the future ecosystem, including \CRANpkg{future.callr}, \CRANpkg{future.mirai}, and HPC schedulers via \CRANpkg{future.batchtools}.
See the package vignettes of \CRANpkg{futurize} for more details and examples on these packages.

\hypertarget{domain-specific-examples}{%
\subsection{Domain-specific examples}\label{domain-specific-examples}}

\hypertarget{bootstrap-with-the-boot-package}{%
\subsubsection{\texorpdfstring{Bootstrap with the `boot' package }{Bootstrap with the `boot' package }}\label{bootstrap-with-the-boot-package}}

The \CRANpkg{boot} package is one of the ``recommended'' R packages, which means it is officially endorsed by the R Core Team, well maintained, of high stability, and by default installed together with R. The package generates bootstrap samples and provides statistical methods around them. Its core function, \texttt{boot()}, produces bootstrap samples of a statistic applied to data. Given its resampling nature, its algorithm is a great candidate for parallelization. The \texttt{boot()} function has built-in mechanisms for parallelization via different combinations of arguments \texttt{parallel}, \texttt{ncpus}, and \texttt{cl}. For example, to parallelize with four workers, we can use:

\begin{verbatim}
library(boot)
cl <- parallel::makeCluster(4)

# Draw 999 samples of the population of 49 large U.S. cities in
# 1920 ('x') and 1930 ('u'). For each sample, calculate the ratio
# of mean-1930 over mean-1920 populations
ratio <- function(pop, w) sum(w * pop$x) / sum(w * pop$u)
b <- boot(bigcity, statistic = ratio, R = 999, stype = "w",
          parallel = "snow", ncpus = length(cl), cl = cl)
\end{verbatim}

Only by inspecting the source code of \texttt{boot()}, do we find that argument \texttt{ncpus} must be strictly greater than one in order for \texttt{parallel\ =\ "snow"} and \texttt{cl\ =\ cl} to have an effect, meaning the default \texttt{ncpus\ =\ getOption("boot.ncpus",\ 1L)} may not be sufficient\footnote{From code inspection of \texttt{boot()}, we find that any \texttt{ncpus} value greater than one is sufficient and gives the same result, i.e., we can also use a constant value, e.g., \texttt{ncpus\ =\ 2L}.}. This adds an extra threshold for a developer who wishes to parallelize the code. The \CRANpkg{futurize} package simplifies this by hiding such details as:

\begin{verbatim}
b <- boot(bigcity, statistic = ratio, R = 999, stype = "w") |> futurize()
\end{verbatim}

\hypertarget{mixed-effects-models-with-the-lme4-package}{%
\subsubsection{\texorpdfstring{Mixed-effects models with the `lme4' package }{Mixed-effects models with the `lme4' package }}\label{mixed-effects-models-with-the-lme4-package}}

The \CRANpkg{lme4} package fits mixed-effects
models. Like \texttt{boot::boot()}, its \texttt{allFit()} function has built-in
mechanisms for parallelization via different combinations of arguments
\texttt{parallel}, \texttt{ncpus}, and \texttt{cl}. Similarly to \texttt{boot::boot()}, all three
arguments need to be set properly in order for \texttt{allFit()} to
parallelize via a PSOCK \pkg{parallel} cluster. The \CRANpkg{futurize}
package abstracts away such details and allows us to parallelize by
piping to \texttt{futurize()}, e.g.

\begin{verbatim}
library(lme4)

# Fit a generalized linear mixed-effects model (GLMM) on data for
# contagious bovine pleuropneumonia (CBPP) comprising 56 observations
# of variables 'incidence', 'size', and 'period'
gm <- glmer(cbind(incidence, size - incidence) ~ period + (1 | herd),
            data = cbpp, family = binomial)
gm_all <- allFit(gm) |> futurize()
\end{verbatim}

\hypertarget{cross-validation-with-the-glmnet-package}{%
\subsubsection{\texorpdfstring{Cross-validation with the `glmnet' package }{Cross-validation with the `glmnet' package }}\label{cross-validation-with-the-glmnet-package}}

The \CRANpkg{glmnet} package uses a highly optimized pathwise coordinate descent algorithm to efficiently compute the entire regularization path for penalized generalized linear models (Lasso, Ridge, Elastic Net). Its \texttt{cv.glmnet()} function performs cross-validation, which can be done in parallel by setting argument \texttt{parallel\ =\ TRUE}, setting up parallel workers and registering a corresponding \CRANpkg{foreach} adapter. The \CRANpkg{futurize} package simplifies this as:

\begin{verbatim}
library(glmnet)

# Simulate 1,000 observations with 100 continuous variables
n <- 1000
p <- 100
x <- matrix(rnorm(n * p), nrow = n, ncol = p)

# Simulate 1,000 continuous response variables
y <- rnorm(n)

# Fit GLM via cross validation
cv <- cv.glmnet(x, y) |> futurize()
\end{verbatim}

\hypertarget{classification-and-regression-training-with-the-caret-package}{%
\subsubsection{\texorpdfstring{Classification and regression training with the `caret' package }{Classification and regression training with the `caret' package }}\label{classification-and-regression-training-with-the-caret-package}}

The \CRANpkg{caret} package provides a rich set of machine-learning tools with a unified API \citep{Kuhn_2008}. The \texttt{train()} function is used to fit models and supports parallel processing by setting argument \texttt{parallel\ =\ TRUE}, setting up parallel workers and registering a corresponding \CRANpkg{foreach} adapter. The \CRANpkg{futurize} package simplifies this as:

\begin{verbatim}
library(caret)

# Fit a random-forest model to the 'iris' data, which has 150 observations
# with five variables, using 100-fold cross validation
ctrl <- trainControl(method = "cv", number = 100)
model <- train(Species ~ ., data = iris, model = "rf", trControl = ctrl) |> futurize()
\end{verbatim}

\hypertarget{other-domain-specific-packages}{%
\subsection{Other domain-specific packages}\label{other-domain-specific-packages}}

The \CRANpkg{mgcv} package \citep{WoodS_2011} is one of the other ``recommended'' packages in R. It provides methods for fitting Generalized Additive Models (GAMs), e.g., the \texttt{bam()} function can be used to fit GAMs for massive datasets (``Big Additive Models'') with many thousand of observations. It supports parallel processing by setting up a \pkg{parallel} cluster and passing it as argument \texttt{cluster}, which is abstracted away by \CRANpkg{futurize}.
The \CRANpkg{tm} package \citep{CRAN:tm} provides a variety of text-mining methods. For example, the map-reduce function \texttt{tm\_map()} can be used to transform a corpus of text and \CRANpkg{futurize} can be used to parallelize it, hiding away package-specific details such as \texttt{tm\_parLapply\_engine()}.
See the package vignettes of \CRANpkg{futurize} for more details and examples on these packages.

\hypertarget{backend-flexibility}{%
\subsection{Backend flexibility}\label{backend-flexibility}}

A key advantage of the future ecosystem approach is backend independence. The same \CRANpkg{futurize} code works across all compliant future backends, enabling concurrent execution that may be realized in parallel, remotely, or sequentially depending on backend. This means that code using \CRANpkg{futurize} can automatically take advantage of any new future backends that become available in the future. Here are some examples of popular future backends being used:

\begin{verbatim}
# Sequential processing (default)
plan(sequential)

# Local PSOCK-based parallel processing using as many workers
# as there are CPU cores available to the current R session
plan(multisession)

# Local PSOCK-based parallel processing (four workers)
plan(multisession, workers = 4)

# Local fork-based parallel processing (Unix/macOS)
plan(multicore, workers = 4)

# Local callr-based parallel processing
plan(future.callr::callr, workers = 4)

# Local mirai-based parallel processing
plan(future.mirai::mirai_multisession, workers = 4)

# Ad-hoc cluster on local computers
plan(cluster, workers = c("n1", "n1", "n2"))

# Ad-hoc cluster on local and remote computers
plan(cluster, workers = c("n1", "n2", "server.example.org"))

# Slurm cluster via batchtools
plan(future.batchtools::batchtools_slurm)
\end{verbatim}

When the number of parallel workers is not specified, it will default to whatever the system allows for\footnote{See \texttt{help("availableCores",\ package\ =\ "parallelly")} \citep{CRAN:parallelly} for which system settings the future ecosystem respects, e.g., Linux Control Groups (``CGroups'') settings \citep{Menage_etal_2008}, HPC scheduler allocations, and a variety of environment variables. This lowers the risk for your R code overusing CPU resources, which is essential on multi-user and multi-service systems.}.

\hypertarget{familiar-behavior-of-stdout-and-condition-handling}{%
\subsection{Familiar behavior of stdout and condition handling}\label{familiar-behavior-of-stdout-and-condition-handling}}

Being able to see and capture output from \texttt{cat()}, \texttt{message()}, etc., and to handle conditions, is something most users take for granted when running R code, but yet, few parallel frameworks support it, if at all. Part of the core design of the future ecosystem is that output produced on parallel workers is relayed as-is in the parent R session. This includes stdout output from functions such as \texttt{cat()}, \texttt{print()}, and \texttt{str()}, and output produced via R's condition framework from functions such as \texttt{message()}, \texttt{warning()}, and \texttt{stop()}. This behavior transfers automatically when using \texttt{futurize()}, e.g.

\begin{verbatim}
library(purrr)
library(futurize)
plan(future.mirai::mirai_multisession)

ys <- 1:4 |> map_dbl(\(x) {
  message("x = ", x)
  sqrt(x)
}) |> futurize()
#> x = 1
#> x = 2
#> x = 3
#> x = 4

ys
#> [1] 1.000000 1.414214 1.732051 2.000000
\end{verbatim}

Output and conditions are relayed as-is in the sense that we can capture and handle them as we do when running code sequentially. For example, we can suppress above messages as:

\begin{verbatim}
ys <- 1:4 |> map_dbl(\(x) {
  message("x = ", x)
  sqrt(x)
}) |> suppressMessages() |> futurize()

ys
#> [1] 1.000000 1.414214 1.732051 2.000000
\end{verbatim}

\hypertarget{progress-reporting}{%
\subsection{Progress reporting}\label{progress-reporting}}

With the \CRANpkg{progressr} package \citep{CRAN:progressr}, progress can be reported from parallelized computations in a near-live fashion:

\begin{verbatim}
library(progressr)
handlers(global = TRUE)

xs <- 1:100
ys <- local({
   p <- progressor(along = xs)
   lapply(xs, \(x) {
     p()
     slow_fcn(x)
   })
}) |> futurize()
\end{verbatim}

Progress updates are propagated from the workers back to the main process, in a near-live fashion, where they are relayed to provide feedback during long-running computations. This works because progress is signaled as R conditions that the \CRANpkg{future} package and most future backends relay instantly.

Note also how \texttt{futurize()} unwraps the expression (Section 3.3), where it descends two levels of wrapped expressions, firstly \texttt{local()} and secondly \texttt{\{\ \}}, to identify the map-reduce \texttt{lapply()} call to be futurized.

\hypertarget{discussion}{%
\section{Discussion}\label{discussion}}

\hypertarget{role-in-the-future-ecosystem}{%
\subsection{Role in the future ecosystem}\label{role-in-the-future-ecosystem}}

The \CRANpkg{futurize} package occupies a unique position in the future ecosystem. While packages like \CRANpkg{future.apply}, \CRANpkg{furrr}, and \CRANpkg{doFuture} provide complete parallel implementations of specific APIs, \CRANpkg{futurize} provides a unified entry point that delegates to these packages internally.
This layered architecture offers several benefits:

\begin{enumerate}
\def\labelenumi{\arabic{enumi}.}
\item
  \textbf{Single point of entry}: Developers and users only need to learn the \texttt{futurize()} function to parallelize any supported API.
\item
  \textbf{Unified options}: The \texttt{futurize()} function provides a consistent way to specify parallel options, hiding the different conventions used by the underlying packages.
\item
  \textbf{Implementation sharing}: The underlying packages continue to handle the details of their respective APIs, avoiding code duplication.
\end{enumerate}

With the introduction of \texttt{futurize()}, from the user's perspective, packages like \CRANpkg{future.apply}, \CRANpkg{furrr}, and \CRANpkg{doFuture} become implementation details. These packages do not need to be attached or even be known by the user, other than that they have to be installed. The \CRANpkg{futurize} package handles the seamless translation, ensuring that the only decision left to the user is the choice of the future backend \texttt{plan()}.

\hypertarget{recommendations-for-package-developers}{%
\subsection{Recommendations for package developers}\label{recommendations-for-package-developers}}

Package developers considering \CRANpkg{futurize} should:

\begin{enumerate}
\def\labelenumi{\arabic{enumi}.}
\item
  \textbf{Parallelization litmus test.} Make sure there are no side effects, or inter-element dependencies in the map-reduce call being parallelized. A simple test is to reverse the processing, e.g., \texttt{rev(lapply(rev(xs),\ fcn))} should give identical results as \texttt{lapply(xs,\ fcn)}, with the exception that random numbers are generated in a different order.
\item
  \textbf{Test with a multi-process backend.} Validate that code works correctly when using independent parallel worker processes, e.g., \texttt{plan(multisession)}.
\item
  \textbf{Handle random numbers correctly.} Use \texttt{seed\ =\ TRUE} when the computation involves random-number generation. The future ecosystem detects when RNG is used without declaring \texttt{seed\ =\ TRUE}, resulting in a warning -- it is important to address such warnings.
\item
  \textbf{Let users control the backend.} Do not set \texttt{plan()} within packages; leave this to end-users. If you do\footnote{If you absolutely have to set the future backend inside a function, make sure to do so temporarily, e.g., \texttt{with(plan(multisession),\ local\ =\ TRUE)}. This avoids overriding the future backend that the user might have set and use elsewhere.}, you lock users into a specific future backend, preventing them from taking advantage of others.\footnote{Scanning CRAN, we find several packages that set \texttt{plan(multisession)} internally in functions. This is unfortunate, because those functions cannot take advantage of modern parallelization backends, e.g., \texttt{plan(future.mirai::mirai\_multisession)}, or future backends that emerge in the future.}
\end{enumerate}

\hypertarget{future-work}{%
\subsection{Future work}\label{future-work}}

The release of the \CRANpkg{future} package in 2015 was followed by community interest in ``futurize'' functionality. Early development focused on identifying a ``syntactic-sugar'' syntax, exploring the use of custom infix operators (e.g., \texttt{\%futurize\%}) or integration with the \CRANpkg{magrittr} pipe (\texttt{\%\textgreater{}\%}), as the native R pipe (\texttt{\textbar{}\textgreater{}}) had not yet been introduced.
Another design challenge was whether hosting of futurize transpilers should be centralized or decentralized. A decentralized model is preferred to minimize maintenance burden and maximize flexibility, but it became evident that centralizing them was the best way to get started. However, there is a plan to develop a standardized API to allow third-party packages to implement their own transpilers;

\textbf{Generic futurization support.} A standard for packages to implement their own futurizing transpilers, e.g., an internal, non-exported function \texttt{futurize\_transpiler()}, which then \texttt{futurize()} could look for. This would allow developers to add \texttt{futurize()} support to their packages without requesting an update to the \CRANpkg{futurize} package. Until available, the author is happy to add them to the \CRANpkg{futurize} package.

Other planned directions for future development are:

\textbf{Validation and troubleshooting.} Mechanisms to test that map-reduce calls give identical results with and without parallelization and with in-order and random-order iterations.

\textbf{Observability and profiling.} Mechanisms to profile map-reduce calls with and without \texttt{futurize()}, and across different future backends, to help developers to better understand how much is gained from parallelization, if at all.

\textbf{Structured Concurrency.} In the context of \CRANpkg{futurize}, structured concurrency \citep{Sustrik_2016, ChenYou_2023} is a relatively modern term where the lifetime of concurrent tasks is limited to the map-reduce construct. The \CRANpkg{future} ecosystem already supports propagation of errors and interrupts and cancellation of futures. Sibling futures are, in most aspects, already canceled when errors or interrupts occur. However, this is not yet guaranteed throughout; for example, user interrupts occurring while futures are being set up may not always cancel futures that have already been launched.

\textbf{Simplified progress reporting.} Add a transpiler \texttt{progressify()} to \CRANpkg{progressr} for injection of progress-reporting code in map-reduce calls, e.g., the verbose code example in Section 4.10 becomes \texttt{ys\ \textless{}-\ lapply(xs,\ slow\_fcn)\ \textbar{}\textgreater{}\ progressify()\ \textbar{}\textgreater{}\ futurize()}.

\textbf{Resource specification API.} Standardized methods for specifying what resources a map-reduce call requires, e.g., memory per iteration and function call, file and mount-point access, and minimal package versions. This can then be used by \texttt{futurize()} and the future ecosystem to identify which parallel workers can take on the task, if at all. Resource specifications can also be used to control the number of parallel tasks in case the amount of available memory is a limiting factor.

\hypertarget{summary}{%
\section{Summary}\label{summary}}

The \CRANpkg{futurize} package is a long-planned enhancement to the \CRANpkg{future} ecosystem that provides a unified, pipe-friendly interface for parallelizing sequential map-reduce operations in R. By transpiling expressions to their future ecosystem equivalents, it allows developers to keep familiar sequential code while enabling parallel execution. The package supports a wide range of APIs from base R, \CRANpkg{purrr}, \CRANpkg{crossmap}, \CRANpkg{foreach}, \CRANpkg{plyr}, \BIOpkg{BiocParallel}, and domain-specific packages. It inherits all the benefits from the core future ecosystem, including as-is propagation of standard output, conditions and errors, automatic identification of global variables, and methods for parallel RNG.

Another key benefit of \CRANpkg{futurize} is its unified options interface through \texttt{futurize()}, which abstracts away the different option conventions used by the underlying parallel packages. This means developers and users can work with a single, consistent set of options regardless of which map-reduce API they are using. This also makes it easy to migrate from \texttt{lapply()} to \texttt{purrr::map()}, or vice versa.

Following the future ecosystem philosophy, \CRANpkg{futurize} maintains a strict separation between what to parallelize (controlled by developers through \texttt{futurize()}) and how to parallelize (controlled by end-users through \texttt{plan()}). This separation ensures that code written with \CRANpkg{futurize} works on any parallel backend, from local multi-CPU-core processing to HPC clusters and cloud computing platforms. The simple pattern \texttt{expr\ \textbar{}\textgreater{}\ futurize()} provides a straightforward path to parallelization that scales from single notebooks to supercomputers without code changes.

\hypertarget{acknowledgments}{%
\section{Acknowledgments}\label{acknowledgments}}

I am grateful to the R community and everyone who has contributed to, provided feedback, bug reports and feature requests on the future ecosystem since the beginning more than a decade ago. The development of the \CRANpkg{future} framework has benefited from grant mechanisms of the R Consortium's Infrastructure Steering Committee (ISC) and the Chan Zuckerberg Initiative (CZI) Essential Open-Source Software (EOSS) program.

\bibliography{bengtsson-futurize.bib}

\address{%
Henrik Bengtsson\\
University of California, San Francisco\\%
Department of Epidemiology and Biostatistics\\ San Francisco, CA 94143, USA\\
\textit{ORCiD: \href{https://orcid.org/0000-0002-7579-5165}{0000-0002-7579-5165}}\\%
\href{mailto:henrik.bengtsson@gmail.com}{\nolinkurl{henrik.bengtsson@gmail.com}}%
}
\end{article}

\pagestyle{plain}
\end{document}